\documentclass[graybox]{svmult}

\usepackage{mathptmx}       
\usepackage{helvet}         
\usepackage{courier}        
\usepackage{type1cm}        
\usepackage{makeidx}         
\usepackage{graphicx}        
\usepackage{multicol}        
\usepackage[bottom]{footmisc}

\makeindex             

\begin{document}

\title*{Utilizing Astroinformatics to Maximize the Science Return of the Next Generation Virgo Cluster Survey}
\titlerunning{Astroinformatics and the NGVS}
\author{Nicholas M. Ball}
\institute{Nicholas M. Ball \at National Research Council Herzberg Institute of Astrophysics, 5071 West Saanich Road, Victoria, BC V9E 2E7, Canada. \email{nick.ball@nrc-cnrc.gc.ca}}
%
%
\maketitle

\abstract*{The Next Generation Virgo Cluster Survey is a 104 square degree survey of the Virgo Cluster, carried out using the MegaPrime camera of the Canada-France-Hawaii telescope, from semesters 2009A--2012A. The survey will provide coverage of this nearby dense environment in the universe to unprecedented depth, providing profound insights into galaxy formation and evolution, including definitive measurements of the properties of galaxies in a dense environment in the local universe, such as the luminosity function. The limiting magnitude of the survey is $g_{AB} = 25.7$ ($10\sigma$ point source), and the $2\sigma$ surface brightness limit is $g_{AB} \approx 29~\mathrm{mag~arcsec}^{-2}$. The data volume of the survey (approximately 50 terabytes of images), while large by contemporary astronomical standards, is not intractable. This renders the survey amenable to the methods of astroinformatics. The enormous dynamic range of objects, from the giant elliptical galaxy M87 at $M(B) = -21.6$, to the faintest dwarf ellipticals at $M(B) \approx -6$, combined with photometry in 5 broad bands ($u$* $g$' $r$' $i$' $z$'), and unprecedented depth revealing many previously unseen structures, creates new challenges in object detection and classification. We present results from ongoing work on the survey, including photometric redshifts, Virgo cluster membership, and the implementation of fast data mining algorithms on the infrastructure of the Canadian Astronomy Data Centre, as part of the Canadian Advanced Network for Astronomical Research (CANFAR).}

\abstract{The Next Generation Virgo Cluster Survey is a 104 square degree survey of the Virgo Cluster, carried out using the MegaPrime camera of the Canada-France-Hawaii telescope, from semesters 2009A--2012A. The survey will provide coverage of this nearby dense environment in the universe to unprecedented depth, providing profound insights into galaxy formation and evolution, including definitive measurements of the properties of galaxies in a dense environment in the local universe, such as the luminosity function. The limiting magnitude of the survey is $g_{AB} = 25.7$ ($10\sigma$ point source), and the $2\sigma$ surface brightness limit is $g_{AB} \approx 29~\mathrm{mag~arcsec}^{-2}$. The data volume of the survey (approximately 50 terabytes of images), while large by contemporary astronomical standards, is not intractable. This renders the survey amenable to the methods of astroinformatics. The enormous dynamic range of objects, from the giant elliptical galaxy M87 at $M(B) = -21.6$, to the faintest dwarf ellipticals at $M(B) \approx -6$, combined with photometry in 5 broad bands ($u$* $g$' $r$' $i$' $z$'), and unprecedented depth revealing many previously unseen structures, creates new challenges in object detection and classification. We present results from ongoing work on the survey, including photometric redshifts, Virgo cluster membership, and the implementation of fast data mining algorithms on the infrastructure of the Canadian Astronomy Data Centre, as part of the Canadian Advanced Network for Astronomical Research (CANFAR).}

\section{Introduction}
\label{sec:1}

The Virgo Cluster is the nearest large cluster of galaxies to us, and as such provides a unique laboratory to study the properties of a dense environment in the local universe, and hence gain insights into galaxy formation and evolution. However, the current state-of-the-art optical surveys of this region are the Virgo Cluster Catalogue \cite{binggeli:vcc}, and the Sloan Digital Sky Survey (SDSS) \cite{york:sdss}, both of which are far inferior to the potential science return of applying modern survey instrumentation to a survey focusing on this particular patch of sky.

The Next Generation Virgo Cluster Survey (NGVS)\footnote{\url{https://www.astrosci.ca/NGVS/The\_Next\_Generation\_Virgo\_Cluster\_Survey}}, is a survey of 104 deg$^2$ that completely supersedes all previous optical surveys of this region. Utilizing the capabilities of the MegaCam camera on the 3.6m Canada-France-Hawaii Telescope, the area is being surveyed to a limiting magnitude of $g_{AB} = 25.7$ ($10\sigma$ point source), and is detecting low surface brightness structures to an unprecedented depth of $g_{AB} \approx 29~\mathrm{mag~arcsec}^{-2}$ ($2\sigma$).

The survey will provide revolutionary improvements in measurements for a host of galaxy and other properties out to the cluster virial radius. The main science goals include: (1) A definitive measurement of the faint-end shape of the cluster galaxy luminosity function; (2) the characterization of galaxy scaling relations over a factor $10^7$ in mass; (3) the study of stellar nuclei in galaxies and their connection to supermassive black holes; (4) the connection between the cluster, galaxies and the intracluster medium; and (5) the fossil record of star formation and chemical enrichment in dense environments. Similar to other large modern wide-field surveys, a great deal of further science will be possible, both in the optical, and in combination with the large number of surveys of the region at other wavelengths, from X-ray to radio. Much science is also possible for objects in the foreground of Virgo (e.g., the Kuiper Belt), and in the background (e.g., 2000+ galaxy clusters).

\subsection{The Role of Astroinformatics Within the NGVS}
\label{subsec:1.1}

The survey provides many opportunities to successfully deliver the promised science by utilizing the methods of astroinformatics\footnote{\url{http://www.ivoa.net/cgi-bin/twiki/bin/view/IVOA/IvoaKDDguide}} \cite{ball:ijmpd,borne:datamining}. At 50T, the data size is such that the survey is not in the petascale regime, in which handling the data volume itself is very much the subject of ongoing research, but it is also substantially larger than the data volume for which many of the traditional astronomical analysis tools are designed.

Therefore, \textit{astroinformatics can be usefully employed on the survey to produce improved science results, without becoming the subject of the research itself}.

Examples of survey challenges that can be addressed utilizing astroinformatics as a tool include:

\begin{itemize}

\item{Data distribution: The 50T dataset size is easily handled by the local infrastructure, but proper data access will still require mature database technology: the SDSS is of similar size, and justified extensive database development.}

\item{Data processing: The large area and high resolution of the survey, combined with the large angles subtended by many galaxies, and 5 passbands, provides a rich dataset of photometric and morphological information. Complex modelling, for example, non-axisymmetric S\'ersic plus nuclear profiles, or simulations of objects, is justified, and will require large amounts of processing time.}

\item{Object detection and classification: The objects in Virgo span an unprecedented dynamic range, from the giant elliptical galaxy M87 ($M(B) = -21.6$) to faint dwarf ellipticals at $M(B) \approx -6$. Objects subtend angles from fractions of an arcsecond for point sources, to degree-scale. Many new low surface brightness features resulting from galaxy interactions are being revealed, which have an irregular, extended structure. This creates problems, e.g., SExtractor does not detect the dwarf ellipticals, and large galaxies prevent local background subtraction.}

\item{Virgo Cluster membership: In the absence of spectroscopic data, one must deduce the probability of cluster membership from photometric data. These populate a high-dimensional space of 5 passbands (4 colours), and numerous morphological parameters. Although traditional criteria such as magnitude vs. surface brightness are quite effective, in general, algorithms that naturally deal with a higher dimensional space will be able to take advantage of the extra information provided by the survey (e.g., morphologically similar objects may have different colours).}

\item{Visualization of results: Many objects within the survey subtend large angles, and as such ideally require the visualization of gigabyte-sized FITS images, possibly overlaid. Standard tools have difficulty coping with the 1.6G data files.}

\end{itemize}


\section{NGVS Survey Data}
\label{sec:2}

The final survey will consist of 50T of images supplied as FITS files. The raw survey data consist of images from the 36 CCD mosaic of the MegaCam camera. These files are pre-processed with the CFHT Elixir pipeline, to provide bias-subtraction, flat-fielding and de-darkening. The Elixir files are then processed by two independent survey pipelines, MegaPipe \cite{gwyn:megapipe,gwyn:cfhtlsstacks}, and TERAPIX \cite{bertin:terapix}. MegaPipe is used to produce stacked images of each survey field, by calibrating each CCD exposure to a high photometric and astrometric precision, and combining the images. This produces, for each 1 square degree survey field, a 1.6G FITS file for each of the 5 bands, $u$* $g$' $r$' $i$' $z$'. Catalogues are then produced from these files.

\section{The CANFAR Computing Infrastructure}
\label{sec:3}

The NGVS survey benefits from being sited at the National Research Council Herzberg Institute of Astrophysics (HIA) in Victoria, British Columbia, which is the home of the Canadian Astronomy Data Centre (CADC). One of the largest astronomy data centres in the world, the CADC has long been at the forefront of providing services for the storage, processing, and distribution of large survey datasets. The mixed environment of astronomers and computer specialists provides an ideal setting for applying astroinformatics to produce science results.

The Canadian Advanced Network for Astronomical Research (CANFAR) \cite{gaudet:canfar}, a collaboration between HIA, the Universities of Victoria and British Columbia, and funded by CANARIE, is a project to provide an infrastructure for data-intensive astronomy projects. This saves a project (e.g., a survey collaboration) time, money, and effort compared to developing a processing pipeline from scratch. The aim is to provide those portions of a pipeline that can be usefully supplied in a generic manner, such as access, processing, storage, and distribution of data, without restricting the analysis that can be performed. Standard astronomical software is provided, but the user may install and run code of their own, as desired.

The system works by combining the job scheduling abilities of a batch system with cloud computing resources. Each user creates and operates one or more Virtual Machines (VMs), in the same way as one would manage a desktop machine. The user submits batch processing jobs which identify the VMs on which they could run. Well-tested, available grid technologies are employed, building on existing CADC infrastructure. Services seen by the user are implemented, where possible, using protocols compatible with the International Virtual Observatory Alliance (IVOA) standards. The CANFAR project was developed throughout with six astronomical science projects, including the NGVS, as an integral part of the process.

Components of the CANFAR infrastructure include a Virtual Image manager, which allows users to create and configure VMs, the Condor job scheduler, which gives the user powerful tools for controlling their jobs, the cloud scheduler, which automatically creates and maintains a cluster consisting of multiple instances of the user VMs, cloud functionality via Nimbus, and virtual distributed storage. The latter is provided by VOSpace, the IVOA standard. This is accessible via the command line (interactively, or from a batch job), or an intuitive graphical interface accessed via a web browser. Data within the system may be either public or proprietary.


\subsection{Use of CANFAR for NGVS Science}
\label{subsec:3.1}

The combination of CANFAR, VOSpace, and astroinformatics tools means that full-scale scientific analysis of real astronomical data is now feasible, by astronomers who are not experts in data-intensive computing. It may be achieved within a realistic timescale, by an international collaboration, for terascale datasets. Currently, as well as the original six projects, several others are beginning to use CANFAR for science data processing that is expected to lead to published results. A CANFAR workshop was held at HIA in May 2011\footnote{\url{http://www.astro.uvic.ca/~canfar/canfarw}}.

Within the NGVS, we have recently created a catalogue of 13,884,023 objects from the first 70 deg$^2$ of the survey (Figure 1)\footnote{Although the present catalogue was not created utilizing the current version of CANFAR, such a catalogue is easily recreated there.}, then used CANFAR to run the code of Schlegel et al. \cite{schlegel:sfd} to provide a Galactic extinction correction for each object. Such extinctions are vital, for example, for obtaining accurate photometric redshifts when using template-based codes. We have used CANFAR to generate simulated galaxies to aid in assessing the completeness of the Virgo Cluster luminosity function, and it will similarly provide the ability to fit detailed galaxy profiles to large numbers of real galaxies within the survey. Many further uses for computationally intensive analysis of the survey that might not otherwise be feasible are planned.

\begin{figure}[h]
\centering
\includegraphics[scale=.68]{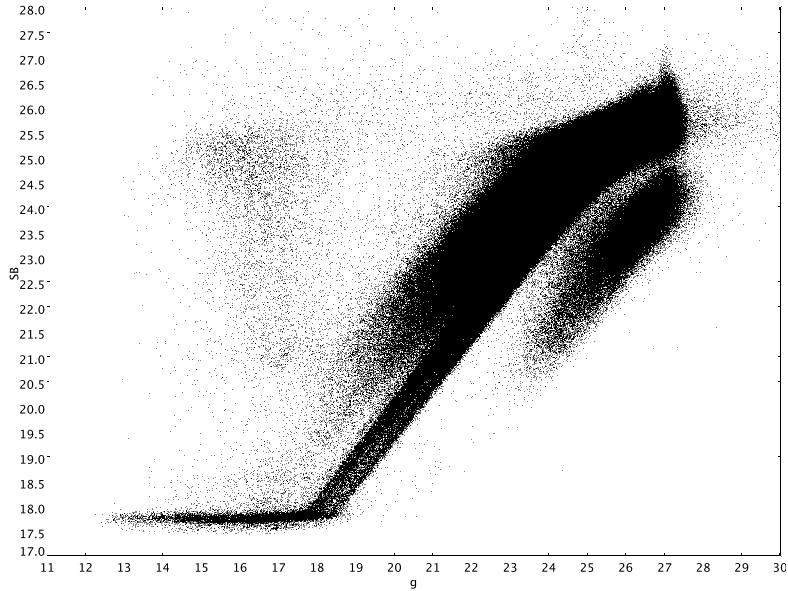}
\caption{Surface brightness versus magnitude for the Next Generation Virgo Cluster Survey, from a SExtractor catalogue of 13,884,023 objects. The plot shows the raw data. Such a catalogue is easily generated using CANFAR compute resources, and the 6.5G FITS file is easily stored and served to survey members via VOSpace. Conversion of the FITS file to column-oriented FITS using STILTS \cite{taylor:stilts} enables the full survey dataset to be plotted in seconds using TOPCAT on a desktop machine. Plotting the full data in this way shows aspects of the raw data that may not be apparent when plotting subsamples, such as the significant spread in the stellar locus, a kink in the distribution at $g \sim 23.5$, and an overdensity of (spurious) objects at $16.75 < g < 17.25$ and $20.75 < SB < 21.25$.}
\label{fig:2}
\end{figure}

\subsection{Fast Data Mining Algorithms}
\label{subsec:3.2}

By extension of the same arguments for providing a hardware infrastructure and standard software tools within CANFAR, we aim to provide a robust set of generic tools which can be used for data analysis. Many modern data mining and machine learning algorithms are ideal for discovering useful patterns in astronomy data, but require time and effort to set up and learn, and, naively implemented, scale in an polynomial way with the number of objects (e.g., $N^2$ or $N^3$). To render the analysis tractable, they must scale no worse than $N\mathrm{log}N$. Such implementations are available, for example, we have confirmed that the proprietary software of the Skytree Corp. scales in this way for typical astronomy data, providing fast runtimes, a wide range of algorithms, and user support that enables its practical use.

Licensing issues may ultimately prevent the deployment of proprietary software on the distributed computing system of CANFAR, so we are also investigating open-source alternatives, such as the Toolkit for Multivariate Data Analysis (TMVA), part of the ROOT system used by the high energy physics community.

\section{Virgo Cluster Membership and Photometric Redshifts}
\label{sec:4}


Whether or not a galaxy is a member of the cluster, or is in the background, is of fundamental importance to many NGVS science goals. For example, a luminosity function measurement for the cluster contaminated by background galaxies will yield a faint-end slope that is too steep. The faint end slope for Virgo remains unconstrained between, for the Schechter function parametrization, $\alpha \approx -1.3$ and $\alpha \approx -1.9$ because of issues like this. This is the difference between the expected value of $\alpha \approx -2$ for a $\Lambda$CDM cosmological model, and heavy suppression of galaxy formation within dark matter haloes, which implies very different physics.

Traditional methods of assigning cluster membership include statistical background subtraction using galaxy number counts, and, for Virgo, the apparent magnitude, surface brightness, luminosity class, and resolution of structure \cite{binggeli:vcc}. However, the NGVS is (a) much deeper than the depths for which spectra can be obtained, and (b) provides 5-band photometry and high resolution morphology. This means that a significantly larger and higher-dimensional sample of information is available, but that it may be non-trivial to extract. For example, although many populations of objects such as low surface brightness dwarf spheroidals must be in Virgo, others, such as compact ellipticals, can be ambiguous, because a small cluster galaxy can mimic a large background galaxy. In general, regions of parameter space will exist where cluster membership assignment must be probabilistic. Thus, the general truth that classifiers should give probabilistic outputs is particularly important in this case.

Results from the NGVS are still in their early stages. However, it is clear that machine learning methods can successfully separate most objects within the cluster from the background. For supervised methods, the usual limitation to the training set regime applies. Here, that is $g < 21$ mag for galaxies, from MMT/Hectospec spectra. Nevertheless, these spectra only cover the central four square degrees of the cluster, so extending this to 104 square degrees represents a considerable advance, although it is likely that environmental variations, e.g., infalling as opposed to {\it in situ} dwarf ellipticals, will create considerable differences in the galaxy populations as a function of cluster environment. One must also, e.g., account for field-to-field variation, although the whole survey is calibrated with reference to the SDSS.

Figure 2 shows the results of one approach to cluster membership: empirical photometric redshifts. At first glance, the use of photometric redshift to assign cluster membership may seem odd, because traditionally, background photometric redshifts have an intrinsic spread of approximately $\Delta z \sim 0.02$, whereas the cutoff for cluster membership is $z < 0.01$. However, the use of the empirical approach with $k$ nearest neighbours (kNN) alleviates this. A support vector machine to classify objects as Virgo or the background gives similar results. Thus, there is currently no strong reason to prefer one machine learning algorithm over the others. Fast and efficient implementations of kNN and SVM were readily available.


Due to poor weather, it is also now unlikely that the full 104 square degrees of the survey will be completed in all 5 bands by the nominal survey completion in semester 2012A. Hence we will obtain a more complex patchwork of results in which some bands are missing for some areas of sky. Methods that can account for missing data, such as a boosted decision tree, may thus provide a suitable approach. This may be of particular importance for regions in which the $u$ band is not available, because the most powerful discriminant of distance, the Balmer break, drops out of $g$ at $z < 0.2$. Hence the ability to include morphological criteria may be critical.

Clearly, there is much potential for refinement of these results. For example, one can assign full probability distributions in redshift space to each galaxy, or one can use unsupervised methods that are not subject to restriction to the spectroscopic regime. These can then be extended from the forced separation into distinct classes of k-means to fuzzy-C means or other methods to provide probabilistic membership. Regions of parameter space that remain ambiguous may be used as motivation for followup spectroscopy. The strength of the astroinformatics approach is that many such methods may be tested, with relative ease, on the entire survey.

\begin{figure}[h]
\centering
\includegraphics[scale=.68]{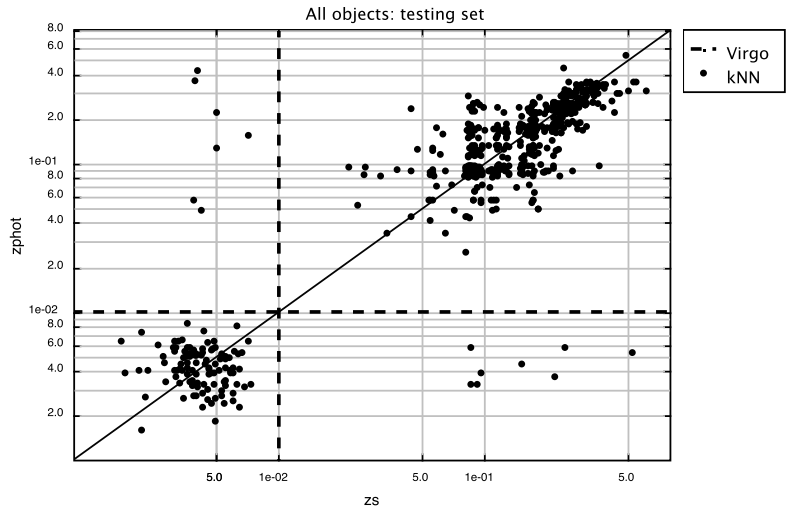}
\caption{Virgo cluster membership: empirical photometric redshift via $k$ nearest neighbours versus spectroscopic redshifts to $g < 21$ mag. The lack of galaxies directly behind Virgo provides a separation such that $z=0.01$ may be taken as a clean cutoff for cluster membership. Objects in the bottom left quadrant are thus correctly assigned Virgo members, and those in the top right are correctly assigned background galaxies. The completeness and efficiency of cluster membership assignment is approximately 90\%. Note that the axes are plotted on a logarithmic scale.}
\label{fig:1}
\end{figure}



\section{Conclusions}
\label{sec:5}

The NGVS is an excellent example of a modern survey for which the methods of astroinformatics are ideally suited. While the data volume (50T) is substantially larger than that for which many of the traditional astronomical analysis tools were designed, it is not so large that it is intractable. This means that astroinformatics can be used to perform science-driven analyses of the survey, maximizing its science potential, without its use being the subject of new research in its own right.

\end{document}